\documentclass{jkas}

\def\year{2024} % publication year
\def\volume{56} % journal volume
\def\issue{1} % journal issue
\def\beginpage{1} % first page of article
\def\received{---} % date paper was received by JKAS
\def\accepted{---} % date of acceptance
\def\published{---} % date of publication
\date{Received \received; Accepted \accepted; Published \published}

\title{%
Dynamical Evolution of Substructured Star Clusters at 3~kpc from the Galactic Center
}

\author[1]{So-Myoung~Park}{0000-0003-1889-325X}
\author[1,$\star$]{Jihye~Shin}{0000-0001-5135-1693}
\author[1]{Sang-Hyun~Chun}{0000-0002-6154-7558}
\author[2]{Simon~P.~Goodwin}{0000-0001-6396-581X}
\author[1]{Kyungwon~Chun}{0000-0001-9544-7021}
\author[3]{Sungsoo~S.~Kim}{0000-0002-5570-2160}
\affil[1]{Korea Astronomy and Space Science Institute, 776 Daedeok-daero, Yuseong-gu, Daejeon 34055, Republic of Korea}
\affil[2]{Department of Physics and Astronomy, University of Sheffield, Sheffield S3 7RH, UK}
\affil[3]{Humanitas College, Kyung Hee University, Yongin, Gyeonggi 17104, Republic of Korea}
\def\corrauthor{%
Jihye~Shin \email{jhshin@kasi.re.kr}
}

\def\runningauthor{%
Park et al.
}

\def\runningtitle{%
YMCs at 3~kpc in the Milky Way
}

\def\keywords{%
methods: numerical --- methods: observational --- Galaxy: open clusters and associations: individual: red supergiant clusters --- stars: kinematics and dynamics
}

\def\abstracttext{%
We investigate the evolution of initial fractal clusters at 3~kpc from the Galactic Center (GC) of the Milky Way and show how red supergiant clusters (RSGCs)-like objects, which are considered to be the result of active star formation in the Scutum complex, can form by 16~Myr.
We find that initial tidal filling and tidal over-filling fractals are shredded by the tidal force, but some substructures can survive as individual subclusters, especially when the initial virial ratio is $\leq$0.5.
These surviving subclusters are weakly mass segregated and show a top-heavy mass function.
This implies the possibility that a single substructured star cluster can evolve into multiple `star clusters'.
}

\begin{document}
\jkashead %% set title, authors, abstract, etc.

%%%%%%%%%%%%%%%%%%%%%%%%%%%%%%%%%%%%%%%%%%%%%%%%%%%%%%%%%%%%%%%%%%%%%
%%% BEGIN MAIN TEXT HERE %%%%%%%%%%%%%%%%%%%%%%%%%%%%%%%%%%%%%%%%%%%%
%%%%%%%%%%%%%%%%%%%%%%%%%%%%%%%%%%%%%%%%%%%%%%%%%%%%%%%%%%%%%%%%%%%%%

\section{Introduction}

Observations have shown that bound star clusters are dense, spherical, and relaxed \citep[e.g.,][]{Lada+2003,Pang+2013}, but gaseous star-forming regions are filamentary, clumpy, and substructured \citep[e.g.,][]{Schneider+2012,Konyves+2015,Lu+2018,Tokuda+2022,Wang+2024}.
Young Massive Clusters (YMCs), which contain a large population of low-mass and high-mass stars, are good laboratories for studying the interstellar medium, star formation, stellar evolution, and cluster dynamics \citep[e.g.,][]{Maschberger+2007,Harris+2009,Kalirai+2010,Park+2018,Krieger+2020}.

At 3~kpc from the GC, six YMCs are observed, containing a large population of red supergiant (RSG) stars, known as red supergiant clusters\footnote{RSGC1, RSGC2, RSGC3, RSGC4 (Alicante 8), RSGC5 (Alicante 7), and RSGC6 (Alicante 10)} \citep[RSGCs; ][]{Clark+2009,Davies+2007,Davies+2008,Gonzalez-Fernandez+2012,Negueruela+2010,Negueruela+2011}.
These YMCs are located in a giant star-forming region, the Scutum complex, where the Scutum-Crux arm and the end of the Galactic Bar meet.
These clusters are separated by projected distances ranging from 31 to 400~pc \citep[][see their figure 4]{Chun+2024}, with RSGC3, RSGC5, and RSGC6 being particularly close to each other, suggesting that they may have originated from a single star-forming region.

\citet{Park+2018} show that a single substructured star cluster can evolve into several star clusters at $R_{\rm g}=30$-$100$~pc, where $R_{\rm g}$ is the Galactocentric radius.
They find that when the size of the initial fractals is smaller than the tidal radius ($r_{\rm t}$), the fractals evolve into a single YMC only for $Q\leq0.5$, where $Q$ is the initial virial ratio.
However, when the initial size of fractal clusters is similar to or larger than the $r_{\rm t}$, the fractals are shredded by the tidal force, but some substructures tend to survive as individual subclusters.
In this paper, we aim to investigate how individual RSGCs-like star clusters, e.g., RSGC3, RSGC5, and RSGC6, can form from a single substructured star cluster at $R_{\rm g}=3$~kpc.

This paper is organized as follows:
In Section~\ref{sec2}, we introduce how we apply the tidal force at $R_{\rm g}=$3~kpc, the initial conditions (ICs), and the method for analyzing surviving star clusters.
Section~\ref{sec3} shows the results of our simulation and in Section~\ref{sec4}, we compare our results with the observations and discuss them.
Section~\ref{sec5} summarizes our results. 

\section{Method}
\label{sec2}

\begin{table}[t!]
    \caption
    {
    Summary of initial conditions for simulations at 3~kpc from the Galactic Center.
    No. is the model number, $\gamma$ is the ratio of the initial fractal radius ($r_{\rm f}$) to the tidal radius ($r_{\rm t}$), $Q$ is the virial ratio, and $D_{\rm f}$ is the fractal dimension.
    We run five different realizations for each case.
    The model numbers in bold are those mainly analyzed in detail in this paper.
    \label{table1}
    }
    \centering
    \begin{tabular}{cccc}
        \toprule
        No. ($\#$) & $\gamma$ & $Q$ & $D_{\rm f}$ \\
        \midrule
        1 & $<$1 & 0.3 & 1.6 \\
        2 & $<$1 & 0.3 & 2.0 \\
        3 & $<$1 & 0.5 & 1.6 \\
        4 & $<$1 & 0.5 & 2.0 \\
        5 & $<$1 & 0.7 & 1.6 \\
        6 & $<$1 & 0.7 & 2.0 \\
        \addlinespace
        {\bf 7} & $\sim$1 & 0.3 & 1.6 \\
        {\bf 8} & $\sim$1 & 0.3 & 2.0 \\
        {\bf 9} & $\sim$1 & 0.5 & 1.6 \\
        10 & $\sim$1 & 0.5 & 2.0 \\
        11 & $\sim$1 & 0.7 & 1.6 \\
        12 & $\sim$1 & 0.7 & 2.0 \\
        \addlinespace
        {\bf 13} & $>$1 & 0.3 & 1.6 \\
        {\bf 14} & $>$1 & 0.3 & 2.0 \\
        {\bf 15} & $>$1 & 0.5 & 1.6 \\
        16 & $>$1 & 0.5 & 2.0 \\
        17 & $>$1 & 0.7 & 1.6 \\
        18 & $>$1 & 0.7 & 2.0 \\
        \bottomrule
    \end{tabular}
\end{table}

We simulate the early dynamics of YMCs using the {\sc nbody6} code \citep{Aarseth1999}, which includes the Galactic (non-truncated) tidal forces \citep{Kim+2000,Park+2018}.
We use clumpy and substructured fractal \citep{Goodwin+2004} ICs at $R_{\rm g}=$3~kpc.

The YMCs evolved for 16~Myr, which is the mean age of the RSGCs \citep[][see their table~A.1.]{Clark+2013}.
We adopt a prescription for stellar evolution consistent with \citet{Kim+1999,Kim+2000}, which calculates the mass loss for individual stars based on a stellar evolution table using linear fitting.
However, to consider the mass loss via stellar winds and explosions, we update the table using the stellar evolution model described in \citet{Hurley+2000}.
Here, we assume Solar metallicity.

\subsection{Tidal field}

We model the Galactic potential using Oort's $A$ and $B$ constants \citep{Oort1927}:
\begin{equation}
	\begin{aligned}
		& A(R_{\rm g})=-\frac{1}{2}R_{\rm g}\frac{{\rm d}\Omega(R_{\rm g})}{{\rm d}R_{\rm G}},\\
        & B(R_{\rm g})=-\Bigg\{\Omega+\frac{1}{2}R_{\rm G}\frac{{\rm d}\Omega(R_{\rm g})}{{\rm d}R_{\rm g}}\Bigg\},
	\end{aligned}
\end{equation}
where $\Omega$ is the orbital angular velocity.
These Oort's constants are calculated using the following assumptions.
Firstly, the profile of the Galactic enclosed mass ($M_{\rm g}$) at 3~kpc from the GC follows a power law \citep{Kim+1999}:
\begin{equation}
	\begin{aligned}
		M_{\rm g}=2.8\times10^{10}~{\rm M}_{\odot}\Bigg(\frac{R_{\rm g}}{3~{\rm kpc}}\Bigg)^{1.3},\\
	\end{aligned}
\end{equation}
where $2.8 \times 10^{10}$~M$_{\odot}$ is the enclosed mass at $R_{\rm g}=3$~kpc.
We calculate the Galactic enclosed mass and the power-law index from \citet[][see their figure~6]{Rodriguez-Fernandez+2008}.

Secondly, the YMCs move in a circular orbit\footnote{The RSGCs are located in a giant star-forming region, the Scutum complex, where the Galactic Bar interacts with the base of the Scutum-Crux arm, which means they are under the influence of the bar potential of the Milky Way.
Nevertheless, we assume that their orbits are nearly circular because we evolve YMCs on a short timescale ($\sim$16~Myr).}, resulting in a time-invariant Galactic potential at a given $R_{\rm g}$:
\begin{equation}
	\label{eq2.3}
	\Omega(R_{\rm g})=\frac{1}{R_{\rm g}}\sqrt{\frac{{\rm G}M_{\rm g}(R_{\rm g})}{R_{\rm g}}},
\end{equation}
where G is a gravitational constant.
We also consider the effective potential to give a more accurate Galactic potential in an acceleration form: 
\begin{equation}
	\frac{{\rm d}^{2}{\boldsymbol{R}_{\rm f}}}{{\rm d}t^{2}}=\frac{2{\rm G}M_{\rm c}{\boldsymbol{R}_{\rm c}}}{{R_{\rm G}}^{3}} - \boldsymbol{\Omega}\ \times\ (\boldsymbol{\Omega}\ \times\ {\boldsymbol{R}_{\rm c}}).
\end{equation}
The first term is the differential gravitational potential and the second term is the centrifugal potential.

\subsection{Initial position and velocity structures}

We choose fractal \citep{Goodwin+2004} ICs for the initial distribution of stars in the YMC to model a complex initial distribution: stars appear to form in clumpy and filamentary substructures \citep{Konyves+2015,Lu+2018,Tokuda+2022,Wang+2024}.
Note that we are not claiming that fractal ICs exactly replicate the process by which giant molecular clouds form stars.
Rather, this method is a more reasonable approximation than the traditional Plummer or King sphere for making clumpy and substructured distributions in the star-forming region.

Fractal distributions are set following \citet{Goodwin+2004}.
A box fractal is constructed in a cube and a sphere is cut from the cube and scaled to the desired total fractal radius ($r_{\rm f}$).
The velocity structure is created by inheriting velocities from a parent star with a small random component that decreases in magnitude as the fractal depth increases, leading to similar velocities among nearby stars and varying velocities among distant stars.
The velocities are then scaled to the desired total $Q$.
See \citet{Goodwin+2004} for full details.
Note that the masses are given randomly, so there is no initial correlation between mass and velocity.

In fractal distributions, the clumpiness is controlled by the fractal dimension ($D_{\rm f}$).
In three dimensions, $D_{\rm f} = 1.6$ is highly substructured, and $D_{\rm f} = 3.0$ indicates a roughly uniform sphere.
In this paper, we consider both $D_{\rm f}=1.6$ (very clumpy) and $D_{\rm f}=2.0$ (moderately clumpy) to investigate how the initial clumpiness affects the results.

\subsection{Cluster mass and IMF}

\begin{table*}[t!]
    \caption
    {
    Results of the models in Table~\ref{table1}.
    We run five different realizations for each case, so $N_{\rm m}$ is the number of realizations with more than 2 surviving subclusters.
    $N_{\rm sv}$ denotes the total number of surviving subclusters.
    In this paper, we use the most representative realization to analyze the results, e.g., the first realization.
    The mean mass of surviving subclusters is represented in $\bar{m}$, and $\alpha$ indicates the median slope of the MF inner half-mass radius ($r_{\rm h}$) for the surviving subclusters.
    The average number of RSG stars in each subcluster is represented in $N_{\rm ms}$, and $Q_{\rm vir}$ means the mean virial ratio of each subcluster.
    }
    \label{table2}
    \centering
    \begin{tabular}{ccccccc}
        \toprule
        No. ($\#$) & $N_{\rm m}$ & $N_{\rm sv}$[min:max] & $\bar{m}$ & $\alpha$ & $N_{\rm ms}$ & $Q_{\rm vir}$ \\
        \midrule
        1 & 1/5 & 1-2 & $\sim$26379~M$_{\odot}$ & $-1.00$ & 279 & 0.33 \\
        2 & 0/5 & 1 & $\sim$26552~M$_{\odot}$ & $-1.03$ & 278 & 0.29 \\
        3 & 0/5 & 1 & $\sim$25265~M$_{\odot}$ & $-1.01$ & 266 & 0.27 \\
        4 & 0/5 & 1 & $\sim$25594~M$_{\odot}$ & $-1.04$ & 270 & 0.26 \\
        5 & 0/5 & 1 & $\sim$22462~M$_{\odot}$ & $-1.02$ & 243 & 0.27 \\
        6 & 0/5 & 1 & $\sim$23248~M$_{\odot}$ & $-1.03$ & 245 & 0.28 \\
        \addlinespace
        {\bf 7} & 5/5 & 2-5 & $\sim$2639~M$_{\odot}$ & $-1.01^{+0.06}_{-0.03}$ & $20^{+25}_{-19}$ & $0.37^{+0.14}_{-0.02}$ \\
        {\bf 8} & 5/5 & 2-5 & $\sim$4321~M$_{\odot}$ & $-1.02^{+0.00}_{-0.04}$ & $21^{+58}_{-5}$ & $0.43^{+0.01}_{-0.00}$ \\
        {\bf 9} & 4/5 & 2-4 & $\sim$2078~M$_{\odot}$ & $-0.95^{+0.02}_{-0.07}$ & $19^{+10}_{-18}$ & $0.39^{+0.16}_{-0.00}$ \\
        10 & 1/5 & 1-2 & $\sim$7566~M$_{\odot}$ & $-1.03$ & 65 & 0.53 \\
        11 & 0/5 & 0-1 & $\sim$8690~M$_{\odot}$ & $-1.01$ & 74 & 0.59 \\
        12 & 0/5 & 0-1 & - & - & - & - \\
        \addlinespace
        {\bf 13} & 5/5 & 3-9 & $\sim$1768~M$_{\odot}$ & $-0.98^{+0.06}_{-0.09}$ & $16^{+21}_{-12}$ & $0.95^{+1.19}_{-0.16}$ \\
        {\bf 14} & 4/5 & 3-10 & $\sim$1213~M$_{\odot}$ & $-1.01^{+0.07}_{-0.04}$ & $9^{+9}_{-5}$ & $0.57^{+0.17}_{-0.02}$ \\
        {\bf 15} & 4/5 & 2-7 & $\sim$1328~M$_{\odot}$ & $-1.00^{+0.09}_{-0.06}$ & $13^{+17}_{-11}$ & $0.84^{+0.53}_{-0.18}$ \\
        16 & 2/5 & 0-3 & $\sim$2804~M$_{\odot}$ & $-1.00$ & 14 & 0.72 \\
        17 & 1/5 & 0-4 & $\sim$1395~M$_{\odot}$ & $-0.97$ & 42 & 1.07 \\
        18 & 0/5 & 0-1 & - & - & - & - \\
        \bottomrule
    \end{tabular}
\end{table*}

We simulate YMCs of mass $\sim$3.0$\times10^{4}~{\rm M}_{\odot}$, which is the initial mean total mass of the RSGCs \citep{Clark+2009,Davies+2007,Davies+2008,Gonzalez-Fernandez+2012,Negueruela+2010,Negueruela+2011}.

We set the total number of stars to $N=46000$ and randomly select the stellar mass from the \citet{Maschberger2013} initial mass function (IMF) within the mass range of stars from 0.01 to 100~M$_{\odot}$.
The Maschberger IMF combines the log-normal approximation for the IMF derived by \citet{Chabrier2003} with the power-law slope of \citet{Salpeter1955} so it is a perfectly standard IMF almost identical to the Chabrier IMF with a much simpler analytic form.
Masses are randomly assigned to each star, so there is no primordial mass segregation.

\subsection{Initial fractal size and internal energy}

\begin{figure*}[t]
    \centering
    \includegraphics[width=150mm]{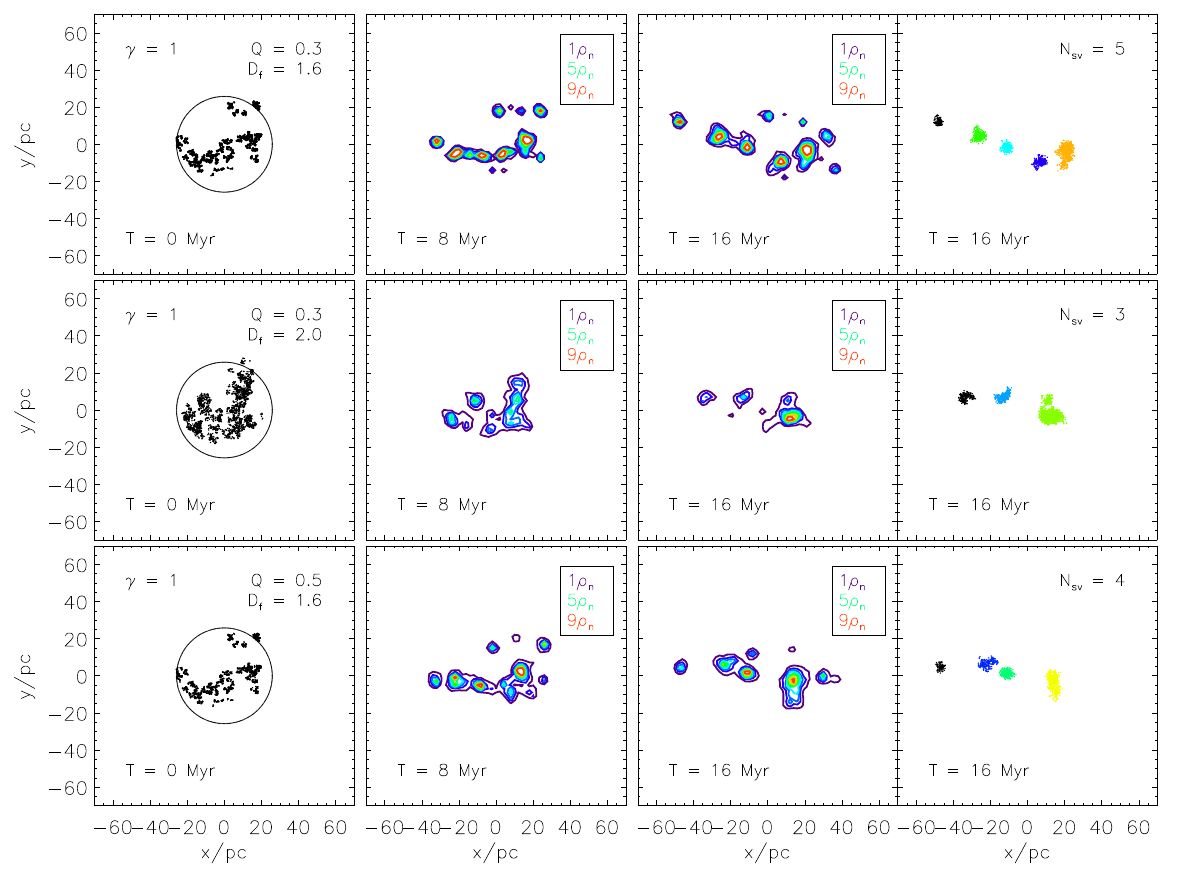}
    \caption
    {
    The evolution of tidal filling fractals ($\gamma\sim$1) at $R_{\rm g}=3$~kpc from the GC.
    The initial fractal distributions are shown in the left panels and the contour map at 8 and 16~Myr is shown in the second and third columns.
    The circle in the first row represents $r_{\rm t}$, and the black dots are stars.
    The colors in the upper right corner of the second and third columns indicate the contour levels.
    The number density inside the half-number radius at 0~Myr is $1\rho_{n}$.
    The surviving subclusters are shown in the right panels after using the FoF algorithm.
    Each color represents the individual subclusters.
    For clarity of the figures, we randomly plot 10\% of stars in the left and right panels.
    \label{fig1}
    }
\end{figure*}

\begin{figure*}[t]
    \centering
    \includegraphics[width=150mm]{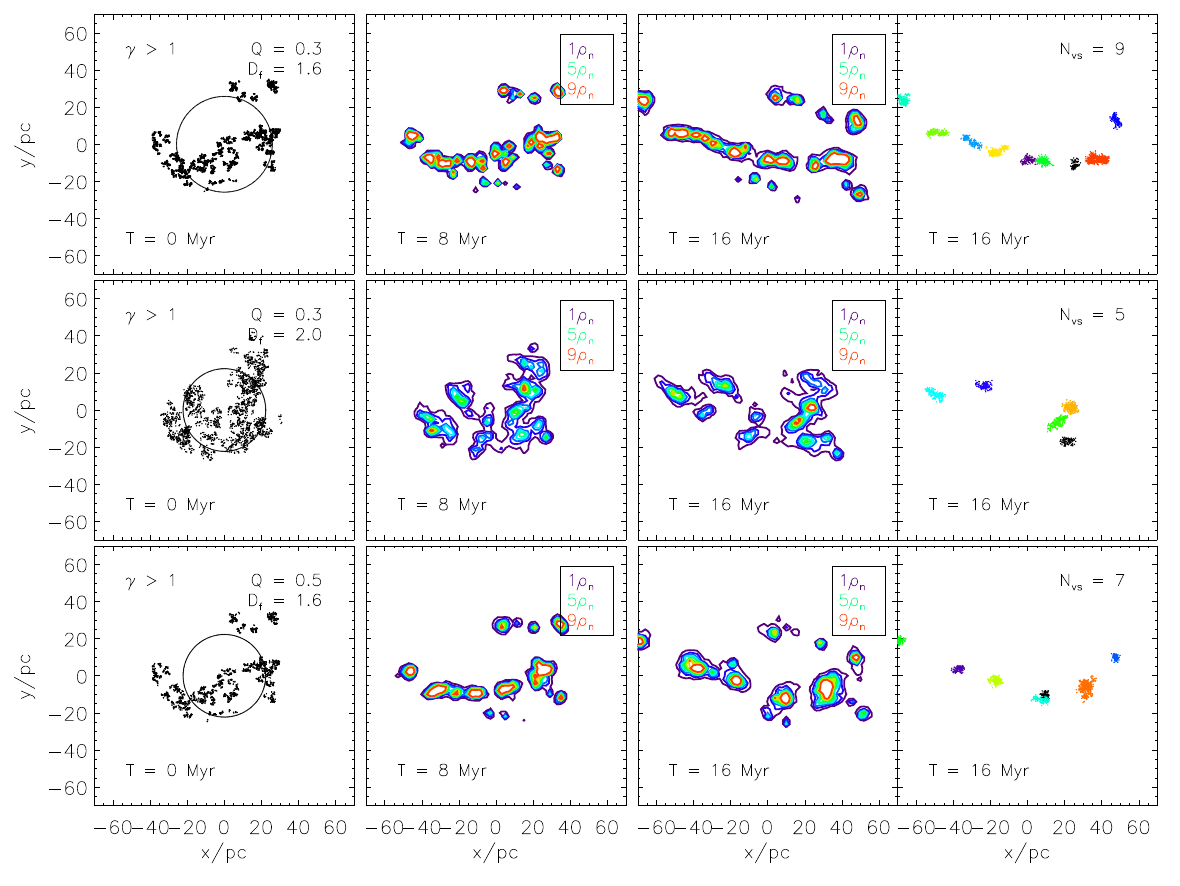}
    \caption
    {
    The evolution of tidal over-filling fractals ($\gamma$>1).
    Symbols are the same as in Figure~\ref{fig1}.
    \label{fig2}.
    }
\end{figure*}

The dynamical evolution of the initial fractal clusters varies depending on the initial fractal radius, $r_{\rm f}$, (the distance from the center of mass to the farthest star) and $Q$ \citep[e.g.,][]{Park+2018}.
We define $r_{\rm t}$ as follows:
\begin{equation}
    r_{\rm t} = \left(\frac{m_{\rm f}}{2M_{\rm g}}\right)^{1/3}R_{\rm g},
\end{equation}
where $m_{\rm f}$ is the mass of the fractal cluster.
Because $r_{\rm t}$ is roughly 26~pc for the initial fractal clusters, we set $r_{\rm f}$ to 13 (tidal under-filling; $\gamma<$1), 26 (tidal filling; $\gamma\sim$1), and 40~pc (tidal over-filling; $\gamma>$1), where the $\gamma$ is a tidal filling ratio, $r_{\rm f}/r_{\rm t}$\footnote{The original tidal filling ratio $\Gamma$ is the ratio of the half-mass radius ($r_{\rm h}$) and the $r_{\rm t}$.
However, the important scale radius of fractals is the total fractal radius ($r_{\rm f}$), which contains all stars, because the $r_{\rm h}$ is poorly defined in clumpy and substructured distributions.
Therefore, we define $\gamma$, using the $r_{\rm f}$ instead of the $r_{\rm h}$.}.

We choose the initial internal energy of our fractal clusters using $Q$, which is the ratio of the kinetic to potential energies of the clusters.
In this paper, we consider three cases: cool ($Q=0.3$), tepid ($Q=0.5$), and hot ($Q=0.7$).
    
\bigskip

A detailed summary of the ICs that we use in this paper is given in Table~\ref{table1}.
Statistically identical fractal ICs can evolve differently because the dynamical evolution of fractals is highly stochastic \citep[e.g,][]{Allison+2010,Park+2020}.
Therefore, for statistical reliability, we use five different realizations for each model.

\subsection{Identification of surviving clusters}
\label{sec2.5}

Initial fractal distributions can be shredded by the tidal force, but some substructures can survive as individual subclusters \citep{Park+2018,Park+2020}.
To identify surviving subclusters, we use the following steps.
First, we calculate the number density ($\rho_{n}$) within the half-mass radius ($r_{\rm h}$) of initial fractal clusters (the left most panels in Figures~\ref{fig1} and \ref{fig2}).
Second, at 16~Myr, regions greater than $\rho_{n}$ are then selected (the middle panels in Figures~\ref{fig1} and \ref{fig2}), and particles within these regions are picked.
Finally, to identify subclusters with picked particles, we use the friends-of-friends (FoF) algorithm \citep{Davis+1985} with a linking length of 0.5~pc, which is slightly larger than the distance between stars in globular clusters (e.g., $\sim$0.3~pc)\footnote{Because the FoF algorithm is quite sensitive, a smaller linking length than 0.5~pc cannot identify subclusters well and the larger linking length than 0.5~pc tends to combine two subclusters into one single cluster.}.
Here, we select FoF clumps with more than 1000 particles and define these clumps as a subcluster.

\subsection{Mass segregation ratio, \texorpdfstring{$\Lambda_{\rm MSR}$}{Lg}}

During the dynamical evolution of star clusters, low-mass and high-mass stars exchange energy to reach the Maxwell-Boltzmann distribution by two-body relaxation.
As a result, high-mass stars lose kinetic energy and move toward the center of the cluster, while low-mass stars gain kinetic energy and move toward the outskirts of the cluster.
This process is called `mass segregation'.
Because initial fractal clusters undergo violent relaxation, they are mass segregated in a few Myrs \citep[e.g.,][]{Allison+2009b,Parker+2014,Park+2018}.
For example, \citet{Allison+2009b} shows that fractal clusters can be dynamically mass segregated down to 2-4~M$_{\odot}$ during the lifetime of the dense core (10-20 crossing times in the core), while their two-body relaxation timescale is longer than the lifetime.
Similarly, even though the two-body relaxation timescale of fractal clusters in this paper is about a few Gyrs, they can be dynamically mass segregated within a few tens of Myrs by violent relaxation.

To quantify the mass segregation of star clusters, we measure the mass segregation ratio, $\Lambda_{\rm MSR}$ \citep{Allison+2009a}.
The value of $\Lambda_{\rm MSR}$ represents the ratio in which a particular set of the $N$ most massive stars is concentrated compared to numerous sets of $N$ random stars of any mass: $\Lambda_{\rm MSR} \sim 1$, $\Lambda_{\rm MSR}$ is $>1$, and $\Lambda_{\rm MSR}<1$ (within `errors') indicates no mass segregation, mass segregation, and inverse mass segregation, respectively.
This method has the advantage of not requiring any assumptions about the location of the `center' of the cluster \citep{Parker+2015}.

\section{Results}
\label{sec3}

We investigate how the initial fractal clusters evolve at $R_{\rm g} = 3$~kpc, examining the effects of the tidal filling ratio ($\gamma=r_{\rm f}/r_{\rm t}$), virial ratio ($Q$), and fractal dimension ($D_{\rm f}$) on their evolution.

We find that tidal under-filling initial fractals evolve into a single cluster regardless of the initial $Q$ and $D_{\rm f}$ (models $\#1$-$\#6$).
The only notable difference is that stars become more dispersed throughout the field as the initial $Q$ increases.

When tidal filling and over-filling fractal clusters are initially cool ($Q=0.3$), they are shredded, but some substructures evolve into surviving subclusters regardless of $D_{\rm f}$ (models $\#7$, $\#8$, $\#13$, and $\#14$).
Fractal clusters that are initially tepid ($Q=0.5$) and very clumpy ($D_{\rm f}=1.6$) also evolve with surviving subclusters during the disruption (models $\#9$ and $\#15$).
Other models, $\#10$-$\#12$ and $\#16$-$\#18$, evolve into a single cluster or are totally disrupted by the tidal force.

In this paper, we are interested in surviving subclusters originating from single fractal cluster.
Among 18 models in Table~\ref{table1}, models of $\#$7, $\#$8, $\#$9, $\#$13, $\#$14, and $\#$15 (the bold model numbers in Table~\ref{table1}) have surviving subclusters at 16~Myr.
For the statistical reliability of the results, we select models with surviving subclusters in at least three out of the five realizations (see $N_{\rm m}$ in Table~\ref{table2}).
These selected models have more than two surviving subclusters at 16 Myr (see $N_{\rm sv}$ in Table~\ref{table2}).

\subsection{Morphology}
\label{sec3.1}

In this section, we examine only the most representative realization of each model among five realizations.
We check that other realizations also show similar results.

Figure~\ref{fig1} shows the initial tidal filling fractal distributions (the left panels) and the contour map at 8 and 16~Myr (the second and third columns).
The right panels show substructures that survive as subclusters, which are the results of the FoF algorithm, at the end of the run (16~Myr).
Similar results are shown at $R_{\rm g}=30$~pc \citep{Park+2018}: tidal filling and cool fractals are shredded by the tidal force, but some substructures can survive as subclusters.
Surviving subclusters at $R_{\rm g}=3$~kpc are spread out to $\sim2$ of their initial $r_{\rm f}$ over 16~Myr.
At $R_{\rm g}=30$~pc, however, surviving subclusters from initial cool fractals are spread out to $\sim$5 times their initial $r_{\rm f}$ over 2~Myr \citep[][see their figure~8]{Park+2018}.
This indicates that surviving subclusters at $R_{\rm g}=3$~kpc are less spread out than those at $R_{\rm g}=30$~pc due to a weaker tidal force.

Initial fractal clusters undergo violent relaxation, erasing their substructure.
The local collapse of each substructure occurs first, followed by the global collapse, and then the cluster attempts to virialize.
In the absence of a tidal field, instead of immediately relaxing into a smooth and virialized distribution, the substructures experience damped oscillation of $Q$ around 0.5 \citep[e.g.,][see their figure~10]{Smith+2011}.
However, if there is a tidal field, certain structures resulting from the local collapse can be taken outside the $r_{\rm t}$ before the global collapse occurs \citep[][see their figures~8 and 13]{Park+2018}.
As a result, some substructures can survive as subclusters even though the fractal clusters evolving in the tidal field are shredded.

In Figure~\ref{fig2}, we show the initial fractal distributions (the left panels), contour maps 8 and 16~Myr (the second and third columns), and final distributions of stars (the right panels), as in Figure~\ref{fig1}, but having increased the tidal filling ratio of $\gamma>1$ (initial tidal over-filling fractal clusters).
Similar to the tidal filling fractal clusters, the tidal over-filling fractal clusters are shredded by the tidal force, but some substructures can survive as subclusters.

As mentioned above, after fractal clusters undergo local collapse, global collapse occurs.
In the case of tidal under-filling fractal clusters, most collapsing substructures are part of the global collapse due to their high density and evolve into a single cluster.
In the case of tidal filling fractal clusters, certain collapsing substructures can merge with others, but before the global collapse occurs, they evolve into individual subclusters due to a tidal force.
For tidal over-filling fractals, however, substructures that experience local collapse are immediately affected by the tidal force before merging with others because most substructures are outside $r_{\rm t}$.
As a result, as $\gamma$ increases, initial fractal clusters evolve into more surviving subclusters at 16~Myr (see $N_{\rm sv}$ in Table~\ref{table2}).
The surviving subclusters of tidal over-filling fractals have mean masses that are half those of the tidal filling fractals (see ${\bar m}$ in Table~\ref{table2}).
We also find that more subclusters can survive for lower $D_{\rm f}$ (see $N_{\rm sv}$ in the top and the middle panels in Figures~\ref{fig1} and \ref{fig2}) and lower $Q$ (see $N_{\rm sv}$ in the top and the bottom panels Figures~\ref{fig1} and \ref{fig2}), because fractal clusters with lower $Q$ and lower $D_{\rm f}$ experience more local collapses.

In the observation, only massive RSG stars can be detected due to a high extinction towards the cluster region.
To investigate the number of RSG stars in each subcluster, we find the stars whose initial mass is 9-40~M$_{\odot}$ \citep[e.g.,][]{Meynet+2000}.
The average number of RSG stars in each cluster is represented as $N_{\rm ms}$ in Table~\ref{table2}.
We find that the value of $N_{\rm ms}$ in tidal filling clusters is slightly larger than the one in tidal over-filling clusters.
Because the average number of RSG stars in RSGCs is $\sim$14, the subclusters from the tidal over-filling clusters have a similar number of RSG stars to the observation.

\subsection{Mass segregation}

\begin{figure}[t]
    \centering
    \includegraphics[width=75mm]{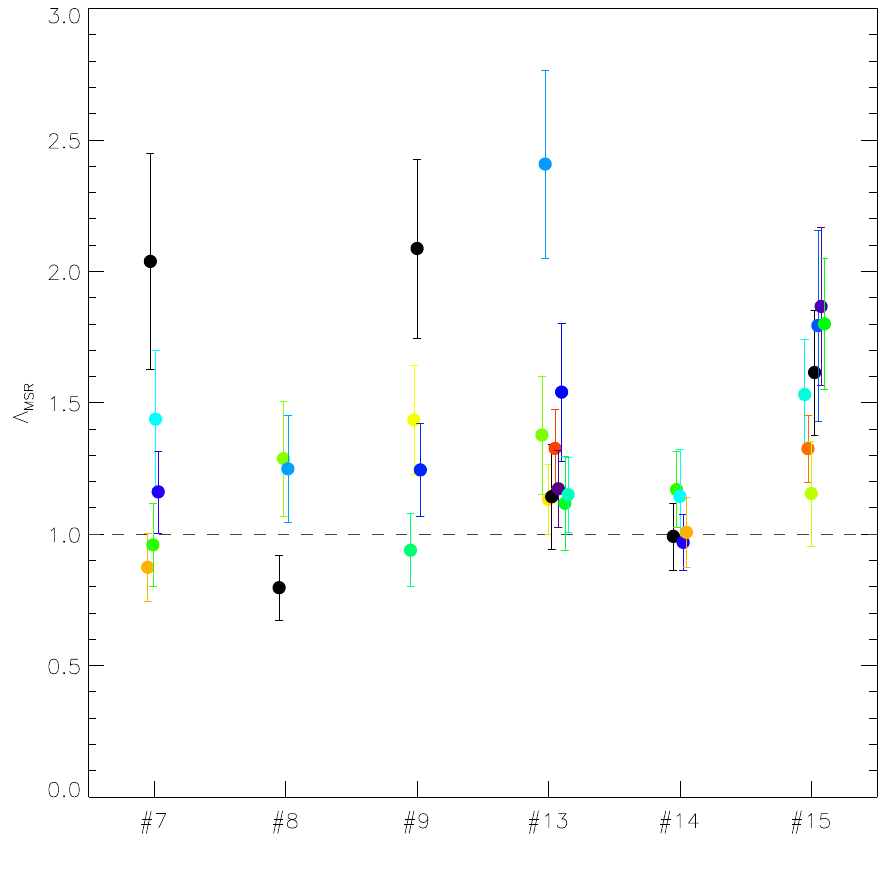}
    \caption
    {
    The mass segregation ratio ($\Lambda_{\rm MSR}$) for the 10 most massive stars of each surviving subcluster in each model at 16~Myr.
    The colored circles are the values of $\Lambda_{\rm MSR}$ of each surviving subcluster, and the vertical bars represent the 1$\sigma$ errors.
    The colors match with the surviving subclusters in Figures~\ref{fig1} and \ref{fig2}.
    The gray horizontal line means no mass segregation.
    \label{fig3}
    }
\end{figure}

Figure~\ref{fig3} shows $\Lambda_{\rm MSR}$ for the 10 most massive stars of each surviving subcluster in each model at 16~Myr.
The typical mass range of the 10 most massive stars and the 10 random stars is $8.6-14.8$~M$_{\odot}$ and $0.1-11.93$~M$_{\odot}$\footnote{Note that the maximum mass of the reference stars can exceed the minimum mass of the 10 most massive stars because their mass varies for subclusters. We checked the mass ranges of the 10 most massive stars and the reference stars do not overlap within any given subcluster.}, respectively.
Regardless of initial $Q$, $D_{\rm f}$, and $\gamma$, the mean value of $\Lambda_{\rm MSR}$ does not show strong mass segregation, although certain subclusters show ($\Lambda_{\rm MSR} > 1.5$).
Interestingly, $\Lambda_{\rm MSR}$ tends to be slightly higher for smaller $D_{\rm f}$, because fractals with smaller $D_{\rm f}$ have more clumpy substructures, leading to more rapid violent relaxation compared to larger $D_{\rm f}$ \citep[e.g.,][]{Allison+2010}.

As mentioned earlier, star clusters are mass segregated, causing low-mass stars to escape from the cluster by two-body relaxation.
In addition, when star clusters evolve into some surviving subclusters, low-mass stars are scattered throughout the field.
This causes surviving subclusters to have more massive stars than low-mass stars.
This means star clusters starting with the canonical IMF can evolve into clusters with top-heavy mass functions (MFs), characterized by an over-abundance of massive stars, leading to a flattened MF slope \citep[e.g.,][]{Park+2020}.

Indeed, surviving subclusters in our models show a flattened MF.
The $\alpha$
values in Table~\ref{table2} represent the median slope of MF of subclusters in the representative realization for stars more massive than 1~M$_{\odot}$ inside $r_{\rm h}$.
We find that all subclusters show $\alpha\sim-1.00$, indicating a significantly flattened MF compared to the classic Salpeter IMF slope of $-1.35$.

\subsection{Stellar density profile}

\begin{figure*}[t]
    \centering
    \includegraphics[width=100mm]{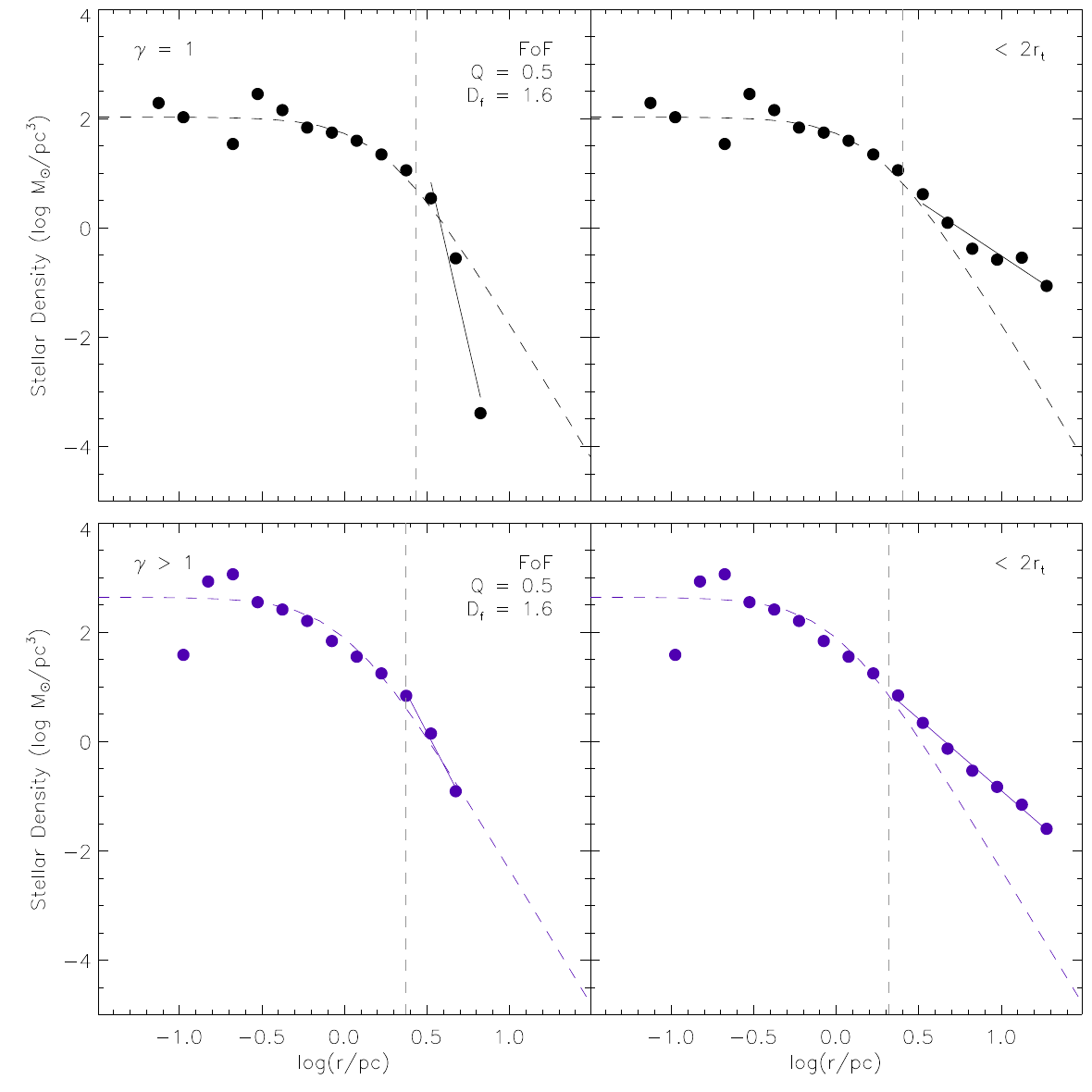}
    \caption
    {
    3D stellar density profiles of initially tepid and tidal filling (top panels) / over-filling (bottom panels) fractal clusters with $D_{\rm f}=1.6$.
    We plot the representative subclusters in each model so that each color corresponds to the color of the subclusters in Figures~\ref{fig1} and \ref{fig2}.
    In the two left panels, we use stars identified as members of subclusters by the FoF algorithm, and in the two right panels, we define stars inside $2r_{\rm t}$ as members of subclusters.
    Colored filled circles show the density profile of each subcluster and colored dashed lines are fitted lines of the Plummer model.
    Dashed gray vertical lines represent $r_{\rm h}$ of each subcluster.
    Colored lines outside $r_{\rm h}$ are linear fit lines.
    \label{fig4}
    }
\end{figure*}

When star clusters evolve in a tidal field, low-mass stars can escape from the cluster due to two-body relaxation, forming tidal tails.
As a result, the density profile of star clusters is represented as a combination of the Plummer (near the center of the cluster) with the shallow power-law profile (outer region of the cluster) due to tidal tails \citep[e.g.,][]{Dalessandro+2015,Park+2018}.

In Figure~\ref{fig4}, we examine the 3-dimensional stellar density profile of initially tepid and tidal filling (top panels) / over-filling (bottom panels) fractal clusters with $D_{\rm f}$=1.6.
Because other models also show similar results, we only plot the representative models.
For clarity, only the representative subcluster in each model is used for the plot.

To investigate the 3D stellar density profile, we identify member stars using the FoF algorithm (see Section~\ref{sec2.5}) and show the results in the two left panels in Figure~\ref{fig4}.
Inside $r_{\rm h}$, the density profile follows the Plummer density profile well.
Beyond $r_{\rm h}$, however, the density profile of tidal filling fractal clusters does not follow the Plummer density profile, rather, it is more like a steep power-law profile.
Or, the density profile of tidal over-filling fractal clusters follows the Plummer profile well beyond the $r_{\rm h}$.
These results are shown regardless of initial $\gamma$, $Q$, and $D_{\rm f}$.
This is because, in our models, most stars are scattered throughout the field due to the tidal force, so the FoF algorithm cannot consider those stars as members of subclusters.

To investigate whether the 3D stellar density profile is affected by how member stars are defined in each subcluster, we identify member stars as those within $2r_{\rm t}$ of each subcluster (two right panels in Figure~\ref{fig4}).
To define $r_{\rm t}$, we adopt the mass of subclusters defined by the FoF algorithm.
Similar to the two left panels, inside $r_{\rm h}$, the stellar density profile follows the Plummer profile well.
However, beyond $r_{\rm h}$, the power-law slope is shallower than that with the FoF algorithm.
This result implies that the stellar density profile is largely affected by how member stars are defined.
It suggests that more accurate member star identification is needed in both simulations and observations, and we note that simulations and observations rarely allocate membership using the same criteria.

\bigskip

In this section, we examine the various dynamical properties of surviving subclusters that have evolved from single fractal clusters with the tidal field at $R_{\rm g}=3$~kpc.
We find that initially cool ($Q = 0.3$) and tepid ($Q=0.5$) fractal clusters are shredded by the tidal force, but some substructures can survive as individual subclusters at 16~Myr.
This result is shown in both initial tidal filling and over-filling fractals.
When fractal clusters are initially very clumpy ($D_{\rm f}=1.6$), they are slightly more mass segregated regardless of the initial virial ratio ($Q$) in both initial tidal filling and over-filling fractals.
In addition, we find that surviving subclusters have a top-heavy mass function.
These results suggest that if clusters at $R_{\rm g} = 3$~kpc are observed to be slightly mass segregated and have a top-heavy MF, they likely form from a single star-forming region.

\section{Comparison with observations}
\label{sec4}

\begin{figure*}[t]
    \centering
    \includegraphics[width=120mm]{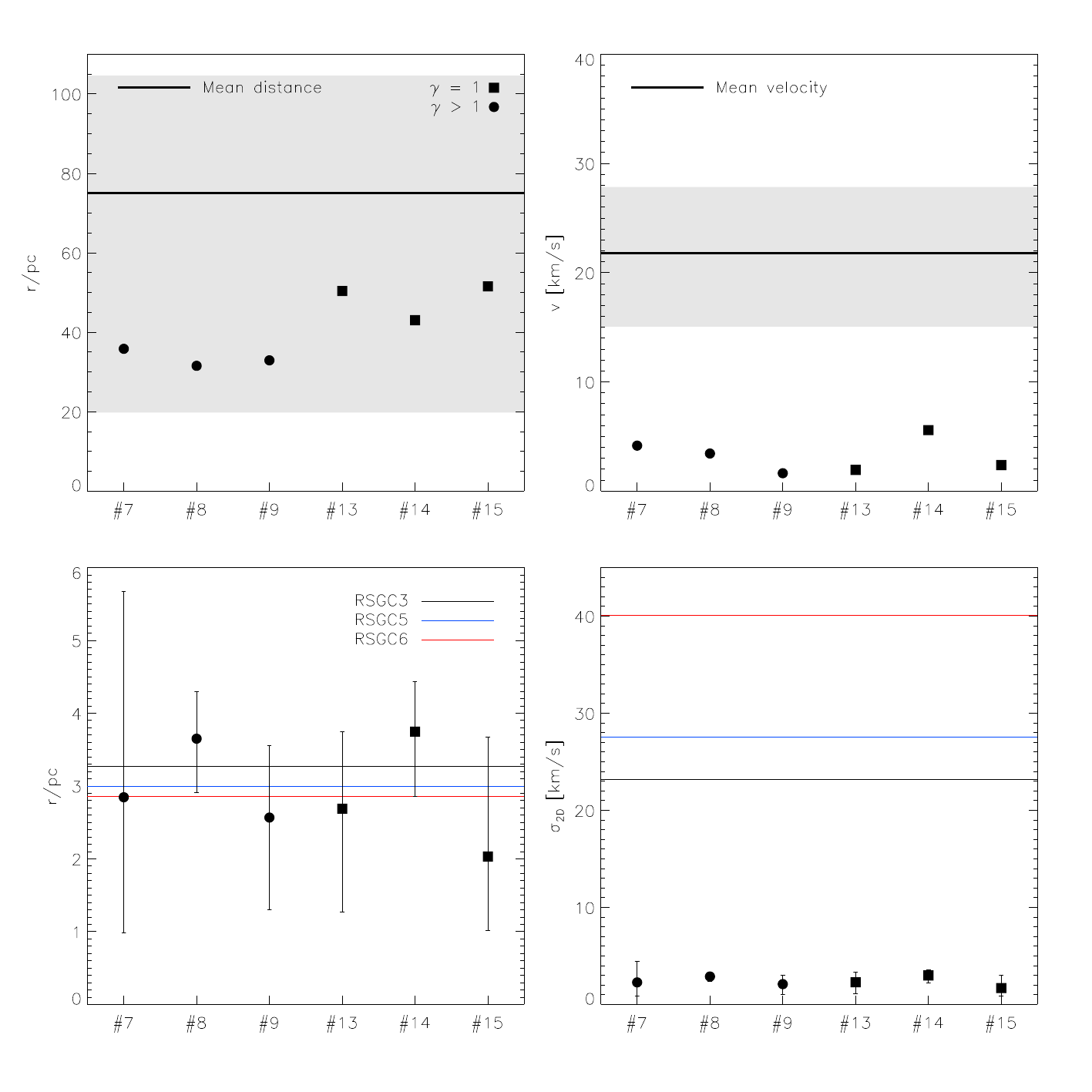}
    \caption
    {
    Mean distance (top left panel) and velocity (top right panel) of all subclusters in each model (black circles and squares).
    The horizontal black lines indicate the mean distance and velocity between RSGC3, RSGC5, and RSGC6, and the gray shaded regions represent the range of the distance and velocity of RSGCs.
    Mean distance (bottom left panel) and 2D velocity dispersion (bottom right panel) of stars in representative subclusters of each model (black circles and squares).
    The colored horizontal lines are the mean distance and velocity between stars in RSGC3, RSGC5, and RSGC6.
    \label{fig5}
    }
\end{figure*}

In Figure~\ref{fig5}, we compare our models with observations using the position and velocity information.
We use the most representative realization in each model.
For the RSGC observations, we note that only the bright and massive RSG stars in clusters are detected due to a high extinction towards the cluster region.
The spatial positions and velocities of individual RSGs in RSGCs were calculated using the distances and radial velocities from \citet{Chun+2024} and the proper motions from Gaia DR3 \citep[][see their Table~1]{Chun+2024}.
We assume the solar position at a Galactocentric distance of 8.2~kpc and a solar velocity of (11.1, 245.04, 7.25)~km/s.
Then, the representative position and velocity for each RSGC are derived by averaging the values of RSGs in RSGCs.
To calculate the velocity errors from the proper motions, we assume the distance error is $\sim$0.8~kpc because RSGCs are located in a range of 5.8-6.6~kpc from the Sun.

The upper left panel shows the mean 2-dimensional distances of the RSGCs (RSGC3, RSGC5, and RSGC6) and all subclusters in each model.
To calculate the mean distance between clusters, we first average the $x-$ and $y-$ positions of stars within each RSGC and subcluster, and then determine their 2D positions.
These averaged 2D positions are defined as the location of each RSGC and subcluster.
The distances between clusters are then calculated based on these defined 2D positions.

The mean distance of tidal filling (models $\#7$-$\#9$) and tidal over-filling fractal clusters (models $\#13$-$\#15$) are about 40$\%$ and 70$\%$ of those observed in RSGCs.
This discrepancy might come from the assumption of the simple Galactic tidal field and the circular orbit at $R_{\rm g}=3$~kpc. 
Thus, we need to consider the realistic Galactic tidal field and the orbit of the Scutum complex, the region where the Galactic Bar and Arm meet.

The subclusters of tidal filling and tidal over-filling fractal clusters are spread out in each model, $\sim$30 and $\sim$50~pc, respectively.
The tidal over-filling fractal clusters are spread out $\sim$1.7 times more because subclusters in tidal over-filling fractals are more affected by the tidal force than those in tidal filling fractals.

The lower left panel shows the mean 2D distance between stars in each model and between RSG stars in each RSGC.
We use the 10 most massive stars in each subcluster to compare with the observations\footnote{In the observation, 14, 8, and 8 RSG stars are detected in RSGC3, RSGC5, and RSGC6, respectively.}.
The mean distance between stars is derived by averaging the distance between stars in both the simulation and observation.
We find that the mean distance between RSGs in the RSGCs is similar to that between stars in the subclusters of each model.
In both tidal and over-filling fractal clusters, stars in moderately clumpy fractals ($\#8$ and $\#14$) are more dispersed than the others.
This is because moderately clumpy fractals undergo weaker local collapse than do very clumpy fractals \citep{Allison+2010}.

In the two panels on the left, the mean distance between RSGCs is larger than that between subclusters in our model, while the mean distance between stars in subclusters and the RSGCs is similar.
This suggests that stars can survive as member stars in RSGCs for 16~Myr even though RSGCs are in the extreme environment of the Scutum complex.

The upper right panel shows the mean 2D relative velocity values of RSGCs (RSGC3, RSGC5, and RSGC6) and those of all subclusters in each model.
To calculate the mean 2D relative velocity distribution, we average 2D relative velocities between clusters in both the simulation and observation after projecting stars into the orbital ($xy$) plane.

The surviving subclusters in our models have similar mean 2D relative velocities, with an average value of $\sim$3.18~km/s, while RSGCs have about 7 times higher mean 2D relative velocities of $\sim$21.81~km/s.
The observed mean 2D relative velocities of the RSGCs is $\sim21.81$~km/s with an error of $\sim6.9$~km/s, suggesting that the discrepancy between the observation and simulation might not come from the observational errors.
Instead, this difference may come from the simplified assumption of the Galactic potential at 3~kpc from the GC in our simulation.
We assume that the fractal clusters are in a circular orbit, so the Galactic potential is time-invariant.
However, RSGCs are located in a violent environment of a large star-forming region known as the Scutum complex, where the base of the Scutum-Crux arm interacts with the Galactic Bar.
Thus, if we apply a more realistic Galactic potential of the Scutum complex, it might explain the discrepancy in mean 2D relative velocities.

A similar trend is shown in the 2D velocity dispersion between stars in subclusters in each model and RSGCs (the bottom right panel).
We use the 10 most massive stars in each subcluster to compare with the observations.
In our simulations, the stars in each subcluster have a low 2D velocity dispersion, which means that the stars are well-bounded, dynamically relaxed, and stable in each subcluster (see also $Q_{\rm vir}$ in Table~\ref{table2}).
However, the RSG stars in each RSGC have higher 2D relative velocities.

The average mass of RSGCs is $\sim10^{4}$~M$_{\odot}$ \citep{Clark+2013}.
For example, the mass of RSGC3 is estimated to be $3\times10^{4}$~M$_{\odot}$, based on the assumption of a Salpeter IMF for the cluster \citep{Figer+2006,Clark+2013}, which may lead to an overestimation of the total mass of the cluster.
Since only high-mass stars are detected in RSGC3, the number of low-mass stars within the cluster remains uncertain.
In our simulations, subclusters show the top-heavy MF (see $\alpha$ in Table~\ref{table2}).
Thus, if one assumes the other MFs, the total mass of RSGC3 could be lower.
Nevertheless, we cannot completely rule out the possibility that the mass of RSGC3 is indeed $3\times10^{4}$~M$_{\odot}$.
This implies that any mass discrepancy might result in an inflated velocity scale, potentially by a factor of around 3, assuming the cluster is in self-gravitating equilibrium.

To further explain this discrepancy, we suggest two possible explanations for the high velocities observed in the RSGCs: 1) the presence of massive binaries, or 2) that we observed RSGCs at a specific moment when they had just recently become unbound.
However, as shown in Table~\ref{table2}, most subclusters remain bound at 16 Myr (see $Q_{\rm vir}$).
Therefore, it is likely that the presence of massive binaries might cause the high velocities observed in the RSGCs \citep[e.g.,][]{Allison+2011}.

Despite the similarities between observational and simulation results, it should be noted that more precise observational data from additional observations are needed for a more accurate comparison.
Only RSGs in RSGC3, RSGC5, and RSGC6 (no low-mass stars) are detected due to high extinction.
As a result, the stellar density profile and the slope of MF are difficult to determine.
\citet{Froebrich+2013} identified possible main-sequence (MS) stars in RSGC3, but spectroscopic follow-up for these MS stars was not conducted.
Therefore, adaptive optics (AO) imaging to detect MS stars in the clusters and subsequent spectroscopic follow-up are highly necessary.
With more accurate proper motions, possibly from Gaia DR4, a more detailed comparison of the MF and velocity distribution with simulation results can be achieved.

\bigskip

Note that we are not claiming that the simulated clusters fully reproduce the observed ones.
We suggest that RSGCs might originate from a single fractal cluster and this is one of the possibilities that RSGCs-like objects could form.
In this paper, we do not consider the scenario in which each RSGC forms from an individual star-forming molecular cloud because the number of RSG stars in each subcluster ($N_{\rm ms}$) of $\#$1-$\#$6 is significantly higher than what is observed (around 10 stars).
Because of the observational limitation (no low-mass stars are detected), we do not exactly know the total mass of each RSGC, the degree of mass segregation, and the slope of the MF.
Instead, since high-mass stars are observed in each RSGC, we can consider $N_{\rm ms}$ as the reliable constraint to compare the simulation with the observation.
We suggest that if we can observe the low-mass stars in each RSGC, we could check the degree of mass segregation, the slope of the MF, and other properties.
These properties could be used as valuable constraints to compare the simulation with the observation, making other scenarios possible.

\section{Summary}
\label{sec5}

We investigate the evolution of substructured star clusters, whose total mass is $\sim$$3.0\times10^{4}$~M$_{\odot}$ with fractal distributions \citep{Goodwin+2004}.
We perform the {\sc nbody6} \citep{Aarseth1999} with the Galactic tidal fields \citep{Kim+2000,Park+2018} at 3~kpc from the Galactic Center and evolve star clusters by 16~Myr, which is the mean age of RSGCs in the Milky Way.
Our results are summarized below.
\begin{enumerate}
    \item When initial tidal filling and over-filling fractal clusters are cool ($Q=0.3$), very clumpy ($D_{\rm f}=1.6$) and moderately clumpy ($D_{\rm f}=2.0$) clusters are shredded by the tidal force, but some substructures can survive as individual subclusters by 16~Myr.
    When the initial tidal filling and over-filling fractal clusters are tepid ($Q=0.5$), very clumpy clusters are destroyed, but some subclusters can form at 16~Myr.
    These surviving subclusters might be the candidate red supergiant clusters (RSGCs).
    
    \item When fractal clusters are very clumpy, surviving subclusters are weakly mass segregated regardless of the initial virial ratio.
    However, there is no mass segregation when fractal clusters are moderately clumpy.
    We find that surviving subclusters have a top-heavy mass function (MF).
    These results imply that clusters at $R_{\rm g}=3$~kpc are observed to be slightly mass segregated and to have a top-heavy MF, they likely form from a single star-forming region in the tidal field.
    
    \item When we compare our models with observations, the distances between clusters and stars are similar.
    However, there is a discrepancy in the 2D relative velocities and velocity dispersions between observations and simulations.
    This discrepancy might come from the simplified Galactic potential in the simulation, the mass discrepancy between simulation and observation, or massive binaries.
\end{enumerate}
Although we suggest that RSGC3, RSGC5, and RSGC6 might form from a single star-forming region, this is one of the possibilities that RSGCs-like objects could form.
Because low-mass stars are not detected due to high extinction, the total mass of each RSGC, the degree of mass segregation, and the slope of the MF are unknown.
These properties could be used as valuable constraints to compare the simulation with the observation, making other scenarios possible.
It suggests that more precise data from additional observations is needed, e.g., the adaptive optics imaging to detect main-sequence stars in RSGCs and subsequent spectroscopic observations.

%%% ACKNOWLEDGMENTS (IF ANY) %%%%%%%%%%%%%%%%%%%%%%%%%%%%%%%%%%%%%%%%
\acknowledgments
This work was supported by the National Research Foundation of Korea (NRF) grant funded by the Korea government (MSIT) (2022M3K3A1093827).
S.-H.C. acknowledges support from the National Research Foundation of Korea (NRF) grant funded by the Korean government (MSIT; NRF-2021R1C1C2003511) and the Korea Astronomy and Space Science Institute under R\&D program (project No. 2024-1-831-00) supervised by the Ministry of Science and ICT.
The work by S.~S.~K. was supported by the Korea Astronomy and Space Science Institute under the R\&D program(Project No. 2023-1-850-09) supervised by the Ministry of Science and ICT.
%%% APPENDICES (IF ANY) %%%%%%%%%%%%%%%%%%%%%%%%%%%%%%%%%%%%%%%%%%%%%
%\appendix
%\section{Appendix Title}

%Some text.

%%% CALL LIST OF REFERENCES (natbib STYLE) %%%%%%%%%%%%%%%%%%%%%%%%%%
\bibliography{RSGCs}

%\begin{thebibliography}{}
%\end{thebibliography}

\end{document}